\documentstyle[aps,twocolumn,twoside]{revtex}
\begin{document}
\draft
\title{Quantum Isomorphic Simulation}
\author{Haiqing Wei and Xin Xue}
\author{\it
Department of ECE, McGill University,
Montreal, Quebec, Canada H3A 2A7
}
\maketitle

\begin{abstract}
\begin{center}
{\bf ABSTRACT}
\end{center}
\vspace{0.2cm}
A new type of quantum simulator is proposed which can simulate any quantum
many-body system in an isomorphic manner. It can actually synthesize a
duplicate of the system to be simulated. The isomorphic simulation has the
great advantage that the inevitable coupling of the simulator to the
environment can be fully exploited in simulating thermodynamic processes.
\end{abstract}
%89.80.+h Computer science and technology
%71.45.-d Collective effects

Simulating quantum many-body systems on a classical computer is hard in the
sense that the simulation takes exponentially long time and large memory as
the size of the system increases [1,2]. The difficulty stems from the fact
that the Hilbert space of the system consists of exponentially many states
as the function of the number of physical variables involved. The efficiency
of simulating a system by another well controlled quantum system (quantum
computer) was conjectured by Feynman [1] and has been justified by Lloyd [2].
Because of the decoherence problem [3], despite the recent advancement of the
quantum computation theory especially on quantum error correction [4,5], it
remains to see whether one can construct a large size quantum computer and
maintain the delicate quantum coherence in order to accomplish meaningful 
computations or simulations, although many efforts have been made to
implement quantum gates or small size quantum computers [6].

Here we propose that a universal quantum isomorphic simulator (IS) can be
constructed by tailoring the many-body interactions among a suitable
physical system. Just for convenience in this paper, the system being
simulated is called the simulatee. One may choose physical systems
consisting of, for example, many isolated quantum wires each with a freely
moving electron inside, or isolated one-dimensional arrays of Josephson
junctions [7] each having a magnetic fluxon freely moving back and forth.
The position of the electron in the wire or the fluxon in the Josephson
junction array provides one continuous degree of freedom which can be
assigned to a continuous variable of the simulatee. For simplicity, we
will use the terms {\it wire} and {\it particle} in the following,
whether they are a quantum wire and an electron or a Josephson junction 
array and a fluxon. Given the Hamiltonian $$H(t)=H_0+H'(t),$$ where
$H_0$ describes the unperturbed system of simulatee and $H'(t)$ is the
external perturbation, one can tailor the many-body interactions among the
system of wires which is called a quantum isomorphic simulator such that
the Hamiltonian describing the simulator is identical (in some sense) to
$H(t)$, the Hamiltonian of the simulatee. This is why it is called an
isomorphic simulator. First note that $H_0-K$ can be decomposed into a sum
of two-body interactions in general, $H_0-K=\sum_{ij}h_{ij}(x_i,x_j)$, where
$x_i,x_j$ are continuous variables which can be simulated by two wires (of
course, discrete variables can also be simulated using wires with double
potential wells or something else), and the external perturbation $H'(t)$
is a sum of one-body potentials, $H'(t)=\sum_kg_k(x_k)$, $K$ is the total
kinetic energy of the simulatee which is naturally mimicked by the inherent
inertial motion of the particles (electrons or fluxons) of the simulator.
One may divide each wire into $N$ parts (discretization with precision
$1/N$), the position-dependent interaction $h_{ij}(x_i,x_j)$ between two
wires can be achieved by drawing $N^2$ connections between the two wires so
that the interaction potential of the $i$th particle at position $x_i$ and
the $j$th particle at position $x_j$ is proportional to $h_{ij}(x_i,x_j)$.
For example, if the wires are Josephson junction arrays and the particles
are magnetic fluxons, one may enclose the magnetic fluxes at each pair of
discrete positions $(x_i,x_j)$ by a superconducting ring so that the
subsystem of the ring and the $i$th and $j$th fluxons has a total
interaction energy $h_{ij}(x_i,x_j)$ when the $i$th fluxon is at position
$x_i$ and the $j$th fluxon is at $x_j$ [8-10].
The one-body potentials $g_k(x_k)$ can be easily realized by applying
position-dependent external fields to the particles. A simulatee with
$M$ continuous variables can be simulated by an IS consisting of $M$
wires, at most $\frac{1}{2}M(M-1)N^2$ connections and no more than $MN$ 
externally applied fields, with $1/N$ discretization precision. The cost
of physical resources is polynomially bounded, and the simulator simulates
the simulatee in an isomorphic manner, {\it i.e.} there is one-to-one
correspondence between the essential Hilbert spaces of the two systems,
with the eigenenergies of two corresponding states proportional to each
other; the two system are governed by the same Hamiltonian (up to a
proportion factor) and undergo the identical dynamics up to a proportion
factor in energy and time {\it etc}. 

To understand more about what isomorphic means, one may compare the IS
with the conventional (quantum) simulator (CS) [2]. On a CS, there is
generally no one-to-one correspondence between the eigenstates of the
simulator and the simulatee with the corresponding energies
proportional to each other. In particular, the ground state
of the simulatee is represented by the ground state of the IS, while the
lowest energy state of the CS can not encode the ground state of the
simulatee because that ground state is not known {\it a priori}.
Although a CS may exploit the inevitable coupling with the environment to
simulate open systems [2], the exploitation is strictly limited. For example,
when simulating the relaxation process of an excited system, a CS should at
least preserve some energy to encode the information for the ground state of
the simulatee. These remaining excitations make the system unstable and
vulnerable to perturbations. By contrast, an IS is not so niggardly, it
happily dissipates all its energy and gets to the lowest energy level
which represents the ground state of the simulatee. This example has
already demonstrated one scheme for an IS to simulate thermodynamic
processes. Since the IS is just a duplicate of the simulatee in the
isomorphic sense, its natural relaxation naturally mimics the relaxation
of the simulatee. This scheme is particularly useful to simulate these
relaxation processes where the actual forms of the external perturbations
are subordinate, what really important are the internal structure of the
simulatee and the statistics of the perturbations. To this extent, an IS
is outstanding in simulating the thermodynamics and dynamics of complex
systems such as biological macromolecules [11] and atomic and molecular
clusters [12]. With a proper Hamiltonian describing the interactions among
smaller pieces of the complex system and some random terms mimicking the
stochastic perturbations, a classical computer may provide some
information, but a full simulation takes
exceedingly long time. By contrast, if provided with the correct Hamiltonian,
an IS can synthesize a duplicate of the complex system, the motion of
the duplicate is identical to the simulatee in the isomorphic sense that
there is one-to-one correspondence between the eigenstates and the
corresponding energies are proportional to each other, the dynamics' of the
two systems are identical up to a constant factor in time scale. For instance,
if provided with the correct Hamiltonian describing the interactions among the
amino acids and their interactions with water, a synthesized {\it protein}
on an IS should {\it fold} itself very fast as the real protein does [11].
On an IS, scientists can test their models and design new complex systems
conveniently. To simulate dynamic processes with well-defined driving
forces on an IS, dissipation should be avoided and one needs to apply
external fields with proper forms to simulate the real driving forces.

In conclusion, the proposed quantum simulator can simulate any quantum
many-body system in an isomorphic manner. The cost of physical resources is
bounded by a polynomial function of the size of the system to be simulated and
the simulation takes a time proportional to the time of the real process. The
great advantage of the isomorphic simulation is that there is a one-to-one
correspondence between the eigenstates of the simulator and the simulatee, the
corresponding energies are proportional to each other, the two systems are
governed by the same equation of motion. In particular, the ground state of
the simulator corresponds to the ground state of the simulatee. Consequently,
an IS is outstanding in simulating natural thermodynamic phenomena. The
inevitable perturbations from the environment can be fully and naturally
exploited when simulating real thermodynamic processes. At present the
well-established technology for superconductive devices is very promising
to implement an IS using Josephson junction arrays [8-10]. The continuing
advancement in fabricating small size devices may eventually make it
possible to implement IS' in terms of semiconductor nanostructures [13].
It is exciting to expect that in the near future quantum isomorphic
simulators are widely used to synthesize and test various complex systems.

\end{document}